\documentclass[final,5p,times,twocolumn]{elsarticle}

\usepackage{graphicx}

\usepackage{amssymb}

\def\lambdabar {\mathchar'26\mkern-10mu\lambda}

\journal{Physics Letters A}

\begin{document}

\begin{frontmatter}


\title{Transition to sub-Planck structures through\\the superposition of $q$-oscillator stationary states}


\author[ugent,ipnas]{E.I. Jafarov\corref{cor1}}
\ead{ejafarov@physics.ab.az}
\author[ugent]{J. Van der Jeugt}
\ead{Joris.VanderJeugt@UGent.be}
\cortext[cor1]{Corresponding author at: Institute of Physics, Azerbaijan National Academy of Sciences, Javid av. 33, AZ-1143 Baku, Azerbaijan}
\address[ugent]{Department of Applied Mathematics and Computer Science, Ghent University\\ Krijgslaan 281-S9, B-9000 Gent, Belgium}
\address[ipnas]{Institute of Physics, Azerbaijan National Academy of Sciences\\ Javid av. 33, AZ-1143 Baku, Azerbaijan}

\begin{abstract}
We investigate the superposition of four different quantum states based on the $q$-oscillator. 
These quantum states are expressed by means of Rogers-Szeg\"o polynomials. 
We show that such a superposition has the properties of the quantum harmonic oscillator when $q\to 1$, 
and those of a compass state with the appearance of chessboard-type interference patterns when $q \to 0$.
\end{abstract}


\end{frontmatter}


\section{Introduction}

The Planck scale plays a central role in quantum mechanics and quantum field theory. 
Following Heisenberg's principle, it was for a long time assumed that phase space structures for a quantum system associated with sub-Planck scales ($\ll\hbar$) do not matter.
In a seminal paper~\cite{zurek-n}, Zurek showed that this assumption is false. He demonstrated that in the phase space of non-local quantum superpositions (or Schr\"odinger cat states) patchy structures on the sub-Planck scale appear. Moreover, he emphasized the physical importance of these sub-Planck scale phenomena for quantum decoherence.

It was known earlier that the generation of `cat states' is possible by the superposition of two minimum-uncertainty Gaussians~\cite{schleich}. 
In the phase space description of such a superposition, interesting structures appear as a result of the interference between `dead' and `alive' states. 
This interference led to the idea of the superposition of four minimum-uncertainty Gaussians as `compass states', where more appealing fine structures appear as a result of the interference between North, South, East and West states~\cite{yurke,hach,lynch,souza}. 
These compass states appear under various names such as orthogonal-even coherent states or pair-cat coherent states.
The compass state, as presented by Zurek~\cite{zurek-n}, is constructed as a superposition of four coherent states. Much of Zurek's attention goes to the study of chessboard-type interference patterns on the sub-Planck scale in the phase space description of this state.
We also refer to other interesting works~\cite{jacquod,toscano,banerji,scott,bhatt,ghosh,austin}, where different aspects of sub-Planck structures were studied by a similar approach.
Note that in all these investigations, the sub-Planck interference phenomena appear only via the appropriate superposition of {\em coherent states}.

All these studies, however, lack information about possible scenarios for the transition from usual quantum mechanical structures to those with sub-Planck lengths. In~\cite{zurek-n}, Zurek notes that the evolution of almost any system will lead into a cat (compass) state, i.e.\ after sufficient time its behaviour will have coherence properties in phase space. 
Then naturally the following question arises: is it possible to consider a scenario when the transition from usual quantum mechanical scales to a system with sub-Planck structures will happen? 
It is clear that in such a scenario, the proposed model should generalize both the `initial' quantum harmonic oscillator and the `final' compass (or cat) states. 
It means that the model should reduce to both the quantum harmonic oscillator and to a compass state at special limits of some parameter. 
Considering that both the quantum harmonic oscillator and the compass state wavefunctions have analytical expressions, 
the proposed superposition should also be some analytic function. 
Taking into account all these requirements, we shall propose in this Letter a model based on the $q$-deformed quantum oscillator.
Quite surprisingly, the superposition of four {\em stationary states} of the $q$-oscillator exhibits, for certain values of the deformation parameter $q$, the sub-Planck interference patterns.
This is, to our knowledge, the first time that such patterns appear in the phase space description without the explicit use of coherent states.

Our Letter is structured as follows: in section 2, we provide basic information about the stationary states of the $q$-deformed oscillator, whose position wavefunctions are expressed by the Rogers-Szeg\"o polynomials and propose the superposition that has the properties of the quantum harmonic oscillator when $q\to 1$, and those of a compass state with the appearance of chessboard-type interference patterns when $q \to 0$. Further properties are discussed in section 3.

\section{The $q$-oscillator and a superposition based on its stationary states}

The stationary states of the $q$-deformed quantum harmonic oscillator in the $x$-representation are given in terms of Rogers-Szeg\"o polynomials~\cite{jafarov, atakishiyev}:
\begin{equation}
\psi_n^{qHO}(x) = c_n\; \mathcal{H}_n(-e^{-2i\lambda hx}|q)\, e^{-\lambda x^2},
\label{qHO}
\end{equation}
where $\lambda$ is given in terms of the mass $m$ and frequency $\omega$: $\lambda  = \frac{{m\omega }}{{2\hbar }}$;
$c_n$ is a normalization constant:
\begin{equation}
\label{c-n}
c_n  = \left( {\frac{{2\lambda }}{\pi }} \right)^{1/4} q^{n/2} \left( {q;q} \right)_n^{-1/2};
\end{equation}
$h$ is a {\em deformation parameter} related to a finite-difference method~\footnote{Note that we use the standard notation $h$ for the step length of the finite-difference method.
This should {\em not} be confused with the Planck constant.
Referring to Planck scales, we always use the reduced Planck constant $\hbar$ in this Letter.} with respect to $x$, and
\begin{equation}
q=e^{-\lambda h^2},\quad 0<q<1\quad(0<h<+\infty). 
\end{equation}

Furthermore, in~(\ref{qHO}) $\mathcal{H}_n$ is the Rogers-Szeg\"o polynomial of the following form~\cite{ismail}:
\begin{equation}
\label{rs-exact}
\mathcal{H}_n \left( { - \tilde x;q} \right) = \sum\limits_{k = 0}^n {\frac{{\left( {q^{ - n} ;q} \right)_k }}{{\left( {q;q} \right)_k }}q^{nk - k^2 /2} \tilde x^k } ,
\end{equation}
where $\left( {a;q} \right)_n$ is the $q$-shifted factorial defined by~\cite{gasper,koekoek}
\begin{equation}
\label{q-shift}
(a;q)_0  = 1,\quad (a;q)_n  = \prod\limits_{k = 0}^{n - 1} (1 - aq^k ),\;(n\geq 1). 
\end{equation}

It is known that in the limit $h\rightarrow 0$, this wavefunction becomes the stationary state of the
ordinary quantum mechanical harmonic oscillator in the $x$-representation~\cite{jafarov}:
\begin{eqnarray}
\label{wave-ho}
&& (-i)^n \psi _n ^{qHO} \left( x \right)\mathop  \to \limits^{h \to 0} \\
&&\psi _n ^{HO} \left( x \right) = \frac{1}{{\sqrt {2^n n!\sqrt {\pi /2\lambda } } }} H_n \left( {\sqrt {2\lambda } \; x } \right) \cdot e^{ - \lambda x^2 },
\nonumber
\end{eqnarray}
where $H_n$ is the usual Hermite polynomial~\cite{koekoek}:

\[
H_n \left( {\tilde x} \right) = n!\sum\limits_{k = 0}^{\left[ {n/2} \right]} {\frac{{\left( { - 1} \right)^k \left( {2\tilde x} \right)^{n - 2k} }}{{k!\left( {n - k} \right)!}}} ,
\]
and $\left[ {n/2} \right]$ denotes the largest integer smaller than or equal to $n/2$.

The stationary states of the $q$-deformed quantum harmonic oscillator can also be determined in the $p$-representation.
This yields an alternative model of the $q$-deformed oscillator. 
The normalized wavefunctions can be expressed through Stieltjes-Wigert polynomials~\cite{jafarov,atakishiyev}, or can be rewritten by means of the Rogers-Szeg\"o polynomials:
\begin{equation}
\tilde \psi_n^{qHO}(x) = c_n\; \mathcal{H}_n(-q^{n-1}e^{-2\lambda hx}|q^{-1})\, e^{-\lambda x^2}.
\label{pHO}
\end{equation}

Again, the limit $h\rightarrow 0$ yields the stationary states of the ordinary quantum harmonic oscillator.

In~\cite{jafarov} it was observed, from the Wigner function of the $q$-deformed oscillator, that the behaviour of the wavefunction 
$\psi_n^{qHO}(x)$ when $h\rightarrow +\infty$ (or $q\rightarrow 0$) is like a coherent state (a Gaussian peak displaced towards $(-\infty,0)$ -- i.e.\ towards the West -- in the $(p,x)$-plane).
This observation leads us to the following superposition proposal:
\begin{eqnarray}
\Psi _{\maltese,n}  \left( x \right) &= \frac{{N_q }}{2}\left[ {e^{in\pi }\psi_n^N \left( x \right) + e^{ 2 in\pi }\psi_n^S \left( x \right)}\right.\nonumber \\ 
& \left.{+ e^{in\pi /2}\psi_n^E \left( x \right) + e^{ 3in\pi /2}\psi_n^W \left( x \right)} \right], \label{compass-gen}
\end{eqnarray}
with the West, East, South and North components
\begin{eqnarray}
 \psi_n^W (x) &=&  c_n e^{-\lambda x^2} \mathcal{H}_n (-e^{-2i\lambda hx}|q)=\psi_n^{qHO}(x), \nonumber\\
 \psi_n^E (x) &=&  c_n e^{-\lambda x^2} \mathcal{H}_n (-e^{2i\lambda hx}|q)=\psi_n^{qHO}(x)|_{h\rightarrow -h}, \nonumber \\ 
 \psi_n^S (x) &=&  c_n e^{-\lambda x^2} \mathcal{H}_n (-q^{n-1} e^{-2\lambda hx} |q^{-1})=\tilde\psi_n^{qHO}(x), \nonumber\\
 \psi_n^N (x) &=&  c_n e^{-\lambda x^2} \mathcal{H}_n (-q^{n-1} e^{2\lambda hx} |q^{-1}) 
 =\tilde\psi_n^{qHO}(x)|_{h\rightarrow -h}.  \label{nsew-exp} \qquad
\end{eqnarray}

The phase factors in~(\ref{compass-gen}) are chosen (according to~(\ref{wave-ho})) in such a way that for the limit $q\rightarrow 1$ ($h\rightarrow 0$), 
all four components in~(\ref{compass-gen}) become a normalized stationary state of the ordinary quantum oscillator.

The normalization constant is found from the following overlap of states~(\ref{compass-gen}):
\begin{equation}
\label{overlap}
\int\limits_{-\infty }^\infty \Psi_{\maltese,m}^* (x) \cdot \Psi_{\maltese,n}(x)dx  = \delta_{mn},
\end{equation}
and it has the following form:
\begin{eqnarray}
\label{n-q}
& N_q  = \left\{ 1 + \frac{q^n}{(q;q)_n}\sum\limits_{k=0}^n \frac{(q^{-n} ;q)_k}{(q;q)_k} \left[ (-1)^n (q^{k};q)_n \right. \right. \nonumber \\ 
 & \left. \left.  + e^{i\frac{n\pi}{2}} (q^{ik};q)_n  + e^{-i\frac{n\pi}{2}} (q^{-ik};q)_n \right]q^{nk}  \right\}^{-1/2} . 
\end{eqnarray}

It is perhaps worth mentioning that for the four components of the wavefunction~(\ref{compass-gen}) one has a Hamiltonian operator, that can be expressed in terms of $q$-generalized
creation and annihilation operators~\cite{jafarov,atakishiyev}. 
These are finite-difference operators and have the following form:
\begin{eqnarray*}
&& b_N^\pm  = \pm \frac{i}{\sqrt{1 - q}} e^{\pm\lambda x^2} ( q^{\pm 1} e^{ 2\lambda hx} e^{\mp h\partial_x} - q^{\frac12} e^{\mp \frac{h}{2}\partial_x})e^{\mp\lambda x^2}, \\ 
&& b_S^\pm  = \mp \frac{i}{\sqrt{1 - q}} e^{\mp\lambda x^2} ( q^{\pm 1} e^{-2\lambda hx} e^{\pm h\partial_x} - q^{\frac12} e^{\pm \frac{h}{2}\partial_x})e^{\pm\lambda x^2}, \\ 
&& b_E^\pm  = \mp \frac{i}{\sqrt{1 - q}} e^{\mp\lambda x^2} ( e^{\pm 2i\lambda hx} - q^{\frac12} e^{ \frac{ih}{2}\partial_x}) e^{\pm\lambda x^2}, \\ 
&& b_W^\pm  = \pm \frac{i}{\sqrt{1 - q}} e^{\mp\lambda x^2} ( e^{\mp 2i\lambda hx} - q^{\frac12} e^{-\frac{ih}{2}\partial_x}) e^{\pm\lambda x^2}.
\end{eqnarray*}
They satisfy the $q$-Heisenberg commutation relation $\left[ b^- ,b^+ \right]_q  = b^- b^+  - qb^+ b^- = 1$ and in the limit $q \rightarrow 1$, all these operators reduce to the well-known quantum harmonic oscillator creation and annihilation operators. 
The components in~(\ref{compass-gen}) satisfy $b^+  b^- \psi_n(x) = [n]_q \psi_n(x)$, where $[n]_q$ is the basic number~\cite{gasper,koekoek}:
\[
\left[ n \right]_q = \frac{{1 - q^n }}{{1 - q}},
\]
which gives the usual number $n$ under the following limit:
\[
\mathop {\lim }\limits_{q \to 1} \frac{{1 - q^n }}{{1 - q}} = n.
\]
Note that the $q$-oscillator, as described here, is an ordinary quantum mechanical system (i.e.\ with the canonical commutation relations). 
The $q$-deformation originates only from the form of the Hamiltonian operator~\cite{jafarov,atakishiyev}.

The main result of the Letter is that $\Psi _{\maltese,n}$ becomes the stationary state of an ordinary quantum oscillator when $h\rightarrow 0$ (a fact already obtained from the earlier limit analysis), and
that it becomes a compass state with sub-Planck structures when $h\rightarrow +\infty$. In order to investigate this last statement, let us compute and study the Wigner distribution function for~(\ref{compass-gen})~\cite{wigner}:
\begin{eqnarray}
\label{wigner-general}
&&W_{\maltese,n} (p,x) = \frac{1}{2\pi\hbar}\\
&&\times \int\limits_{-\infty}^\infty  \Psi_{\maltese,n}^* ( x - \frac{x'}{2})\; \Psi_{\maltese,n} (x + \frac{x'}{2}) e^{-\frac{ipx'}{\hbar}} dx'. \nonumber
\end{eqnarray}

This function can be computed explicitly and consists of $16$ components computed by the combination of $N$, $S$, $E$ and $W$ wavefunctions~(\ref{compass-gen}):
\begin{eqnarray}
&&W_{\maltese,n} (p,x) = W_n^{NN} (p,x) + e^{in\pi } W_n^{NS} (p,x) \nonumber\\ 
&&  + e^{ - in\pi /2} W_n^{NE} (p,x) + e^{in\pi /2} W_n^{NW} (p,x) \nonumber\\ 
&&  + e^{ - in\pi } W_n^{SN} (p,x) + W_n^{SS} (p,x) + e^{ - 3in\pi /2} W_n^{SE} (p,x) \nonumber\\ 
&&  + e^{ - in\pi /2} W_n^{SW} (p,x) + e^{in\pi /2} W_n^{EN} (p,x) \nonumber\\ 
&&  + e^{3in\pi /2} W_n^{ES} (p,x) + W_n^{EE} (p,x) + e^{in\pi } W_n^{EW} (p,x) \nonumber\\ 
&&  + e^{ - in\pi /2} W_n^{WN} (p,x) + e^{in\pi /2} W_n^{WS} (p,x) \nonumber\\ 
&&  + e^{ - in\pi } W_n^{WE} (p,x) + W_n^{WW} (p,x). \label{wig-16}
\end{eqnarray}

For the computation of these components, we refer to~\cite[\S~4]{jafarov}, and just present the final expressions here:
\begin{eqnarray*}
&& W_n^{NN}(p,x) = \frac{N_q^2}{4\pi\hbar}(-1)^n q^{ - \left( {\scriptstyle n \hfill \atop 
  \scriptstyle 2 \hfill} \right)} e^{ - \frac{2}{\hbar \omega }\left( \frac{m\omega^2 x^2 }{2} + \frac{p^2}{2m} \right)} \\ 
&&\times {\;} _3\varphi_2 \left( \begin{array}{*{20}c}
   q^{-n} ,q^n e^{-ia} ,q^n e^{ia^*}  \\
   q,0  \end{array}; q,q  \right) = W_n^{SS} (-p,-x),
\end{eqnarray*}
\begin{eqnarray*}
&&W_n^{NS} (p,x) = \frac{N_q^2}{4\pi\hbar}e^{-\frac{2}{\hbar\omega}\left(\frac{m\omega^2 x^2}{2} + \frac{p^2}{2m} \right)}  \cdot e^{ina} \\ 
&&\times {\;}_3 \varphi_2 \left( \begin{array}{*{20}c}
   q^{-n} ,qe^{-ia} ,e^{ia^*} \\
   q,0  \end{array}; q,q  \right) = W_n^{SN}(-p,-x), 
\end{eqnarray*}
\begin{eqnarray*}
&&W_n^{NE} (p,x) = \frac{N_q^2}{4\pi\hbar}\frac{q^n}{(q;q)_n}e^{-\frac{2}{\hbar\omega}\left(\frac{m\omega^2x^2}{2} + \frac{p^2}{2m}\right)} \\ 
&&\times \sum\limits_{k=0}^n \frac{(q^{-n};q)_k}{(q;q)_k} (q^{ik} e^a;q)_n (q^n e^{ia^*})^k = {W_n^{EN}}^* (p,x), 
\end{eqnarray*}
\begin{eqnarray*}
&&W_n^{NW} (p,x) = \frac{N_q^2}{4\pi\hbar}\frac{q^n}{(q;q)_n}e^{-\frac{2}{\hbar\omega}\left(\frac{m\omega^2x^2}{2} + \frac{p^2}{2m}\right)} \\ 
&&\times \sum\limits_{k=0}^n \frac{(q^{-n};q)_k}{(q;q)_k} (q^{-ik} e^{-a};q)_n (q^n e^{ia^*})^k = {W_n^{WN}}^* (p,x),
\end{eqnarray*}
\begin{eqnarray*}
&&W_n^{SE} (p,x) = \frac{N_q^2}{4\pi\hbar}\frac{q^n}{(q;q)_n}e^{-\frac{2}{\hbar\omega}\left(\frac{m\omega^2x^2}{2} + \frac{p^2}{2m} \right)} \\ 
&& \times \sum\limits_{k=0}^n \frac{(q^{-n};q)_k}{(q;q)_k} (q^{-ik} e^a ;q)_n (q^n e^{-ia^*})^k  = {W_n^{ES}}^* (p,x), 
\end{eqnarray*}
\begin{eqnarray*}
&&W_n^{SW} (p,x) = \frac{N_q^2}{4\pi\hbar}\frac{q^n}{(q;q)_n}e^{-\frac{2}{\hbar\omega}\left( \frac{m\omega^2 x^2}{2} + \frac{p^2}{2m} \right)} \\ 
&& \times \sum\limits_{k=0}^n \frac{(q^{-n};q)_k}{(q;q)_k}(q^{ik} e^{-a} ;q)_n (q^n e^{-ia^*})^k  = {W_n^{WS}}^*(p,x), 
\end{eqnarray*}
\begin{eqnarray*}
&&W_n^{EE} (p,x) = \frac{N_q^2}{4\pi\hbar} (-)^n q^{-\left( {\scriptstyle n \hfill \atop \scriptstyle 2 \hfill} \right)} 
 e^{-\frac{2}{\hbar\omega}\left( \frac{m\omega^2 x^2}{2} + \frac{p^2}{2m} \right)} \\ 
&&\times {\;}_3\varphi_2 \left( \begin{array}{*{20}c}
   q^{-n} ,q^n e^a ,q^n e^{a^*}  \\
   q,0  \end{array}; q,q \right) = W_n^{WW} (-p,-x), 
\end{eqnarray*}
\begin{eqnarray*}
&&W_n^{EW} (p,x) = \frac{N_q^2}{4\pi\hbar}e^{-\frac{2}{\hbar\omega}\left( \frac{m\omega^2 x^2}{2} + \frac{p^2}{2m}\right)} \cdot e^{-na} \\ 
&& \times {\;}_3 \varphi_2 \left( \begin{array}{*{20}c}
   q^{-n} ,qe^a ,e^{a^*}  \\
   q,0  \end{array}; q,q \right) = W_n^{WE} (-p,-x),
\end{eqnarray*}
where, $a=\frac{h}{\hbar} p + 2i\lambda h x$ and ${}_3 \varphi_2(\cdot)$ is the basic hypergeometric series of the following form~\cite{gasper}:
\[
{}_3 \varphi_2 \left( \begin{array}{*{20}c}
   q^{-n} ,a_1 ,a_2   \\
   b_1 ,b_2 \end{array}; q,z \right) = \sum\limits_{k=0}^n \frac{( q^{-n},a_1 ,a_2 ;q)_k }{(b_1 ,b_2 ,q;q)_k }z^k ,
\]
with $(\alpha,\beta,\gamma ;q )_k  \equiv (\alpha ;q)_k (\beta ;q)_k (\gamma ;q)_k$.

\section{Discussions}

\begin{figure*}
\begin{center}
\includegraphics[scale=0.5]{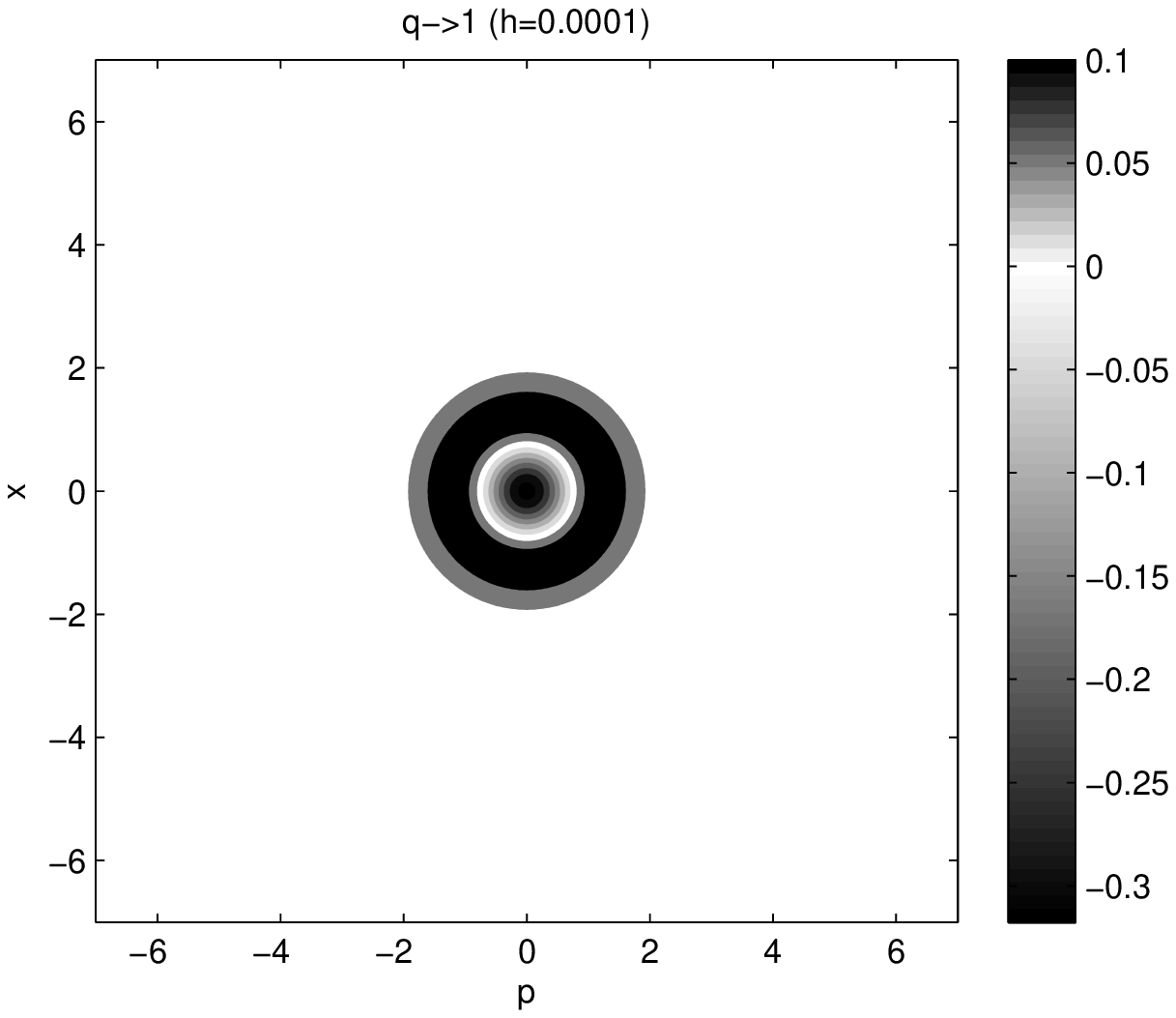}
\includegraphics[scale=0.5]{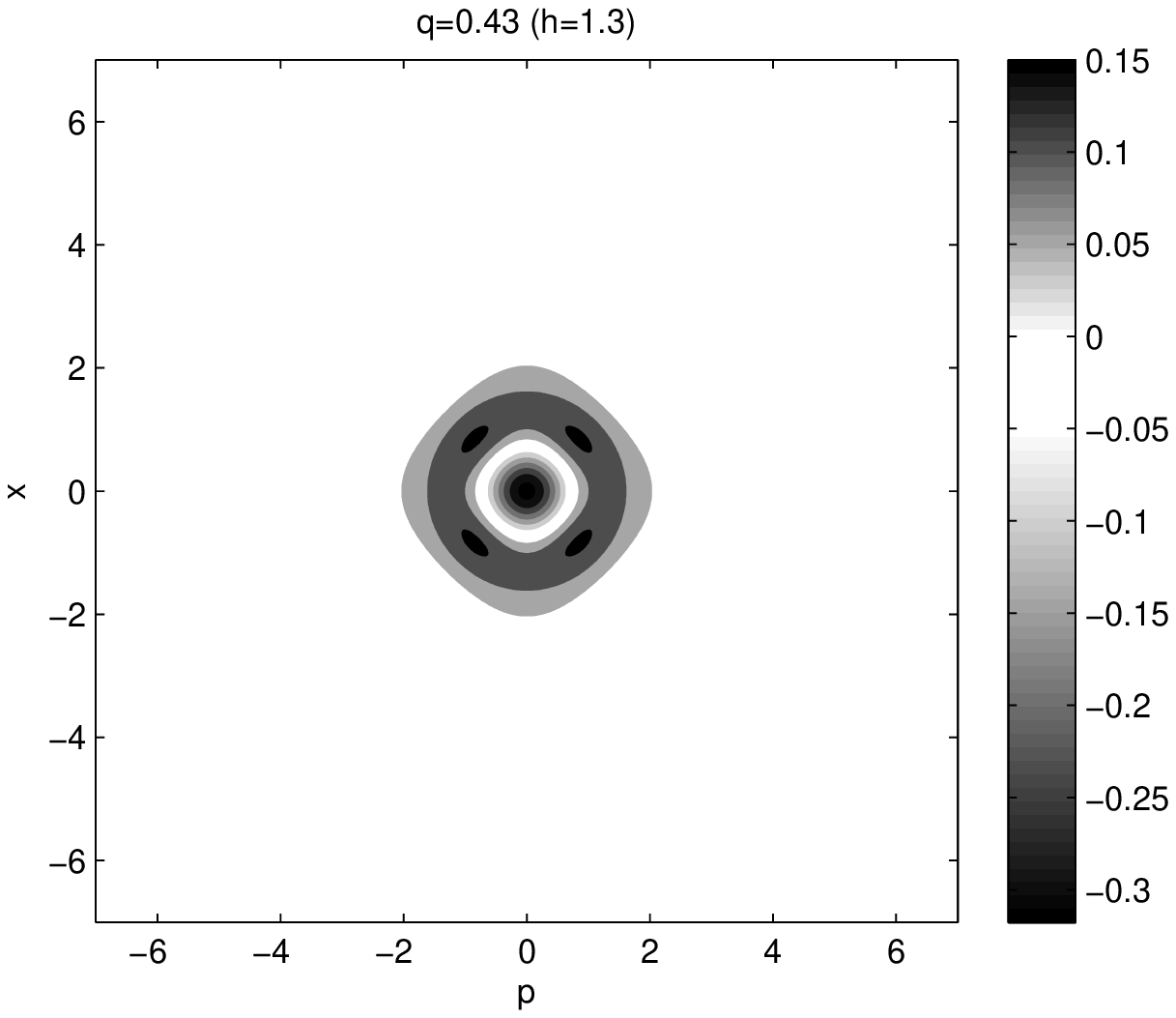}\\
\includegraphics[scale=0.5]{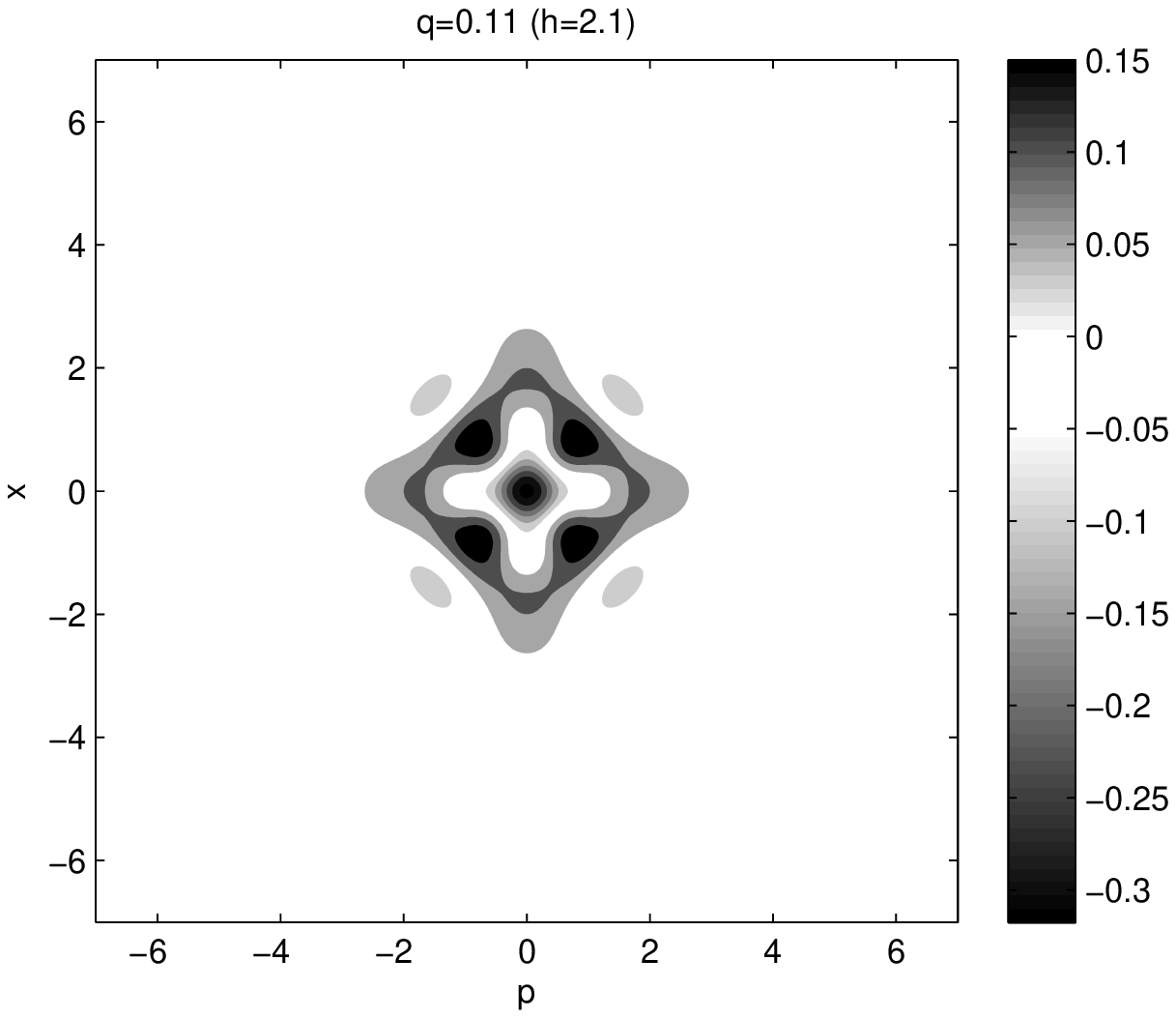}
\includegraphics[scale=0.5]{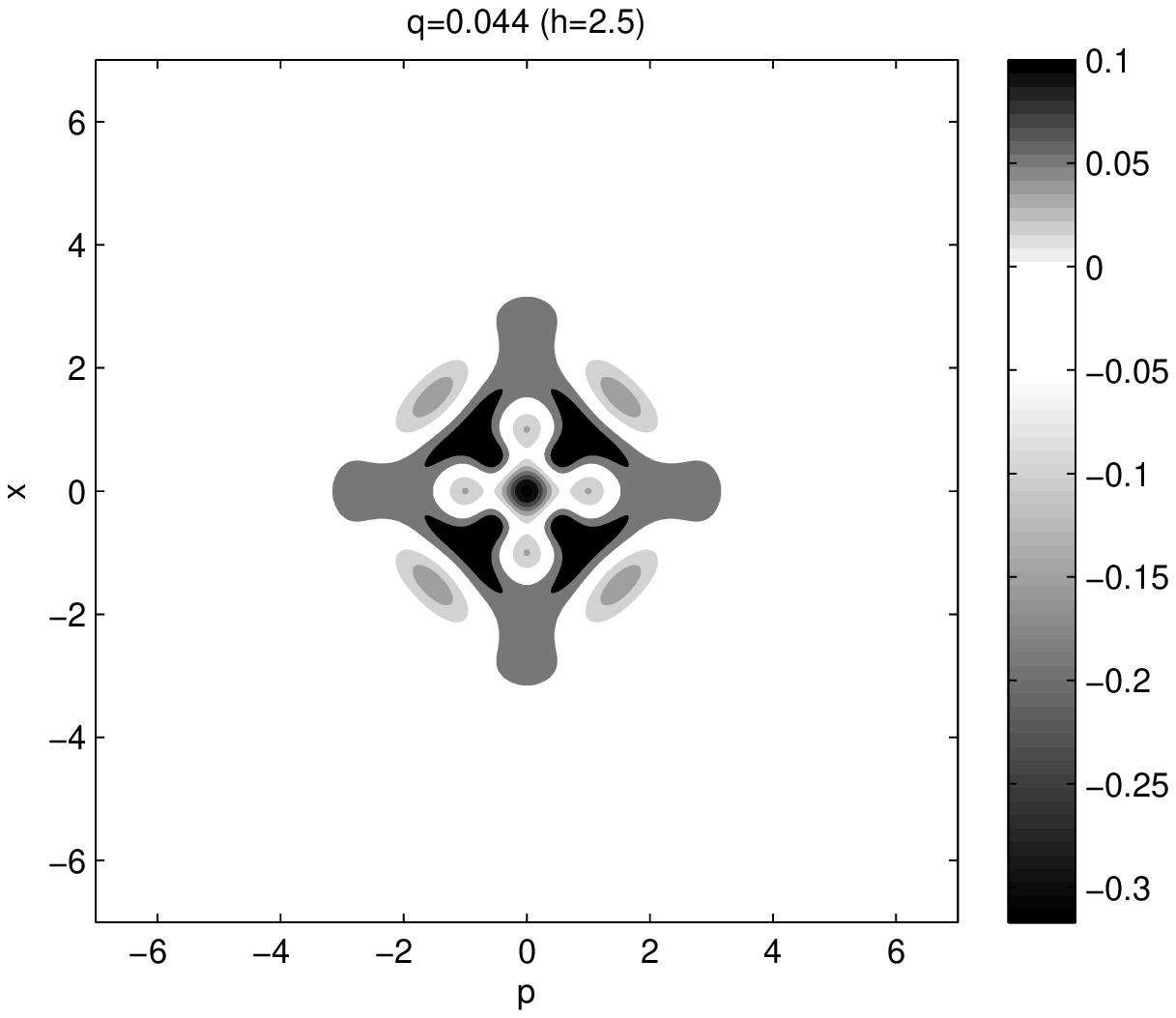}\\
\includegraphics[scale=0.5]{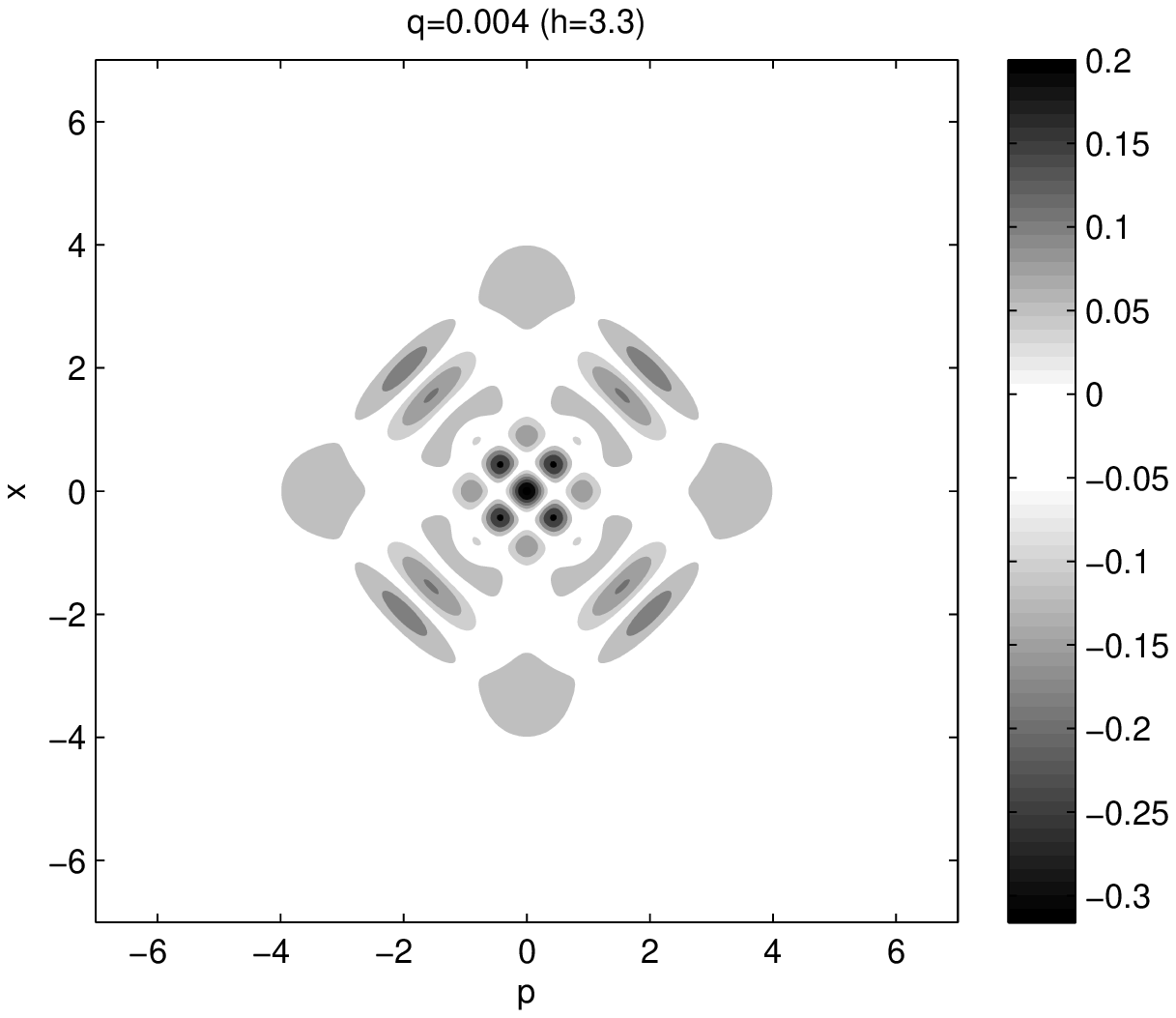}
\includegraphics[scale=0.5]{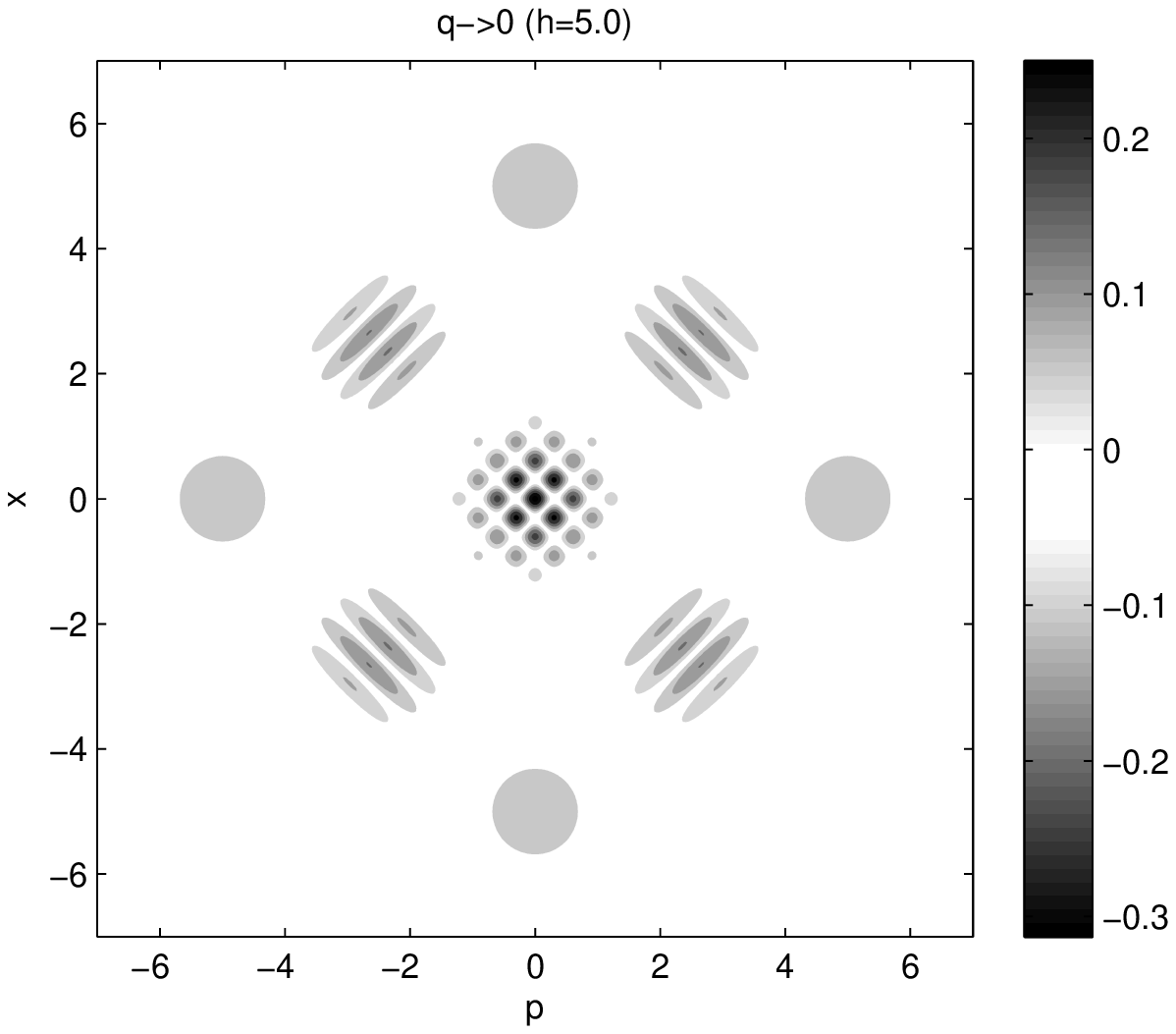}
\end{center}
\caption{\label{fig1}A density plot of the Wigner function of the single photon state ($n=1$) of~(\ref{compass-gen}), for values of $h=0.0001$, $1.3$, $2.1$, $2.5$, $3.3$, $5.0$ and $m=\omega=\hbar=1$. 
The value of $q$ is found through the relation $q=\exp(-h^2/2)$.}
\end{figure*}

In figure~\ref{fig1} we show a density plot of the Wigner function of the single photon state ($n=1$) of~(\ref{compass-gen}). 
For simplicity, we use $m=\omega=\hbar=1$. 
One can see that for values of $h$ close to $0$ (this corresponds to the limit case $q \to 1$) the system under consideration coincides with the non-relativistic quantum harmonic oscillator.
The first signs of sub-Planck structures can be seen in the second plot of figure~\ref{fig1}, which corresponds to the value $h=1.3$ ($q=0.43$). 
By form and behaviour, it can be considered as a scaled-up `square' from the chessboard-type pattern that appears in the compass state superposition. 
The next plot, which corresponds to the value $h=2.1$ ($q=0.11$) contains information about the possible evolution of the system to the compass state, i.e.\ it can be considered as a primitive compass state. 
As can be seen from the next two plots, by increasing the value of $h$ we can observe the formation of four Gaussian-like states directed to the north, south, east and west. 
As one can see from last plot, for values of $h>>0$ ($q \to 0$) they become Gaussians and their peaks are located at a distance equal to $nh$ ($m=\omega=\hbar=1$) from the origin~\cite[(5.3)]{jafarov}.
Note that in figure~\ref{fig1}, since we work in the scale $\hbar=1$, the patchy interference structures that appear are obviously on a sub-Planck scale.

It can be verified that in the Wigner function~(\ref{wig-16}), the four terms $W_n^{NN}(p,x)$, $W_n^{SS}(p,x)$, $W_n^{EE}(p,x)$ and $W_n^{WW}(p,x)$ are responsible for the four Gaussian
peaks directed to North, South, East and West respectively. 
The four sums of the form $e^{-in\pi/2} W_n^{NE}(p,x) +e^{in\pi/2} W_n^{EN}(p,x)$ (and similarly for $NW$, $SE$ and $SW$) are responsible for the interference patterns in the four directions $NE$, $NW$, $SE$ and $SW$.
Finally, the sum $e^{in\pi} W_n^{NS}(p,x) + e^{-in\pi} W_n^{SN}(p,x) + e^{in\pi} W_n^{EW}(p,x) + e^{-in\pi} W_n^{WE} (p,x) $ creates the chessboard-type pattern in the middle.
Analyzing this sum in more detail, one would observe that it is a real function of $p$ and $x$ consisting of the Gaussian $e^{-\frac{2}{\hbar\omega}\left(\frac{m\omega^2x^2}{2} + \frac{p^2}{2m}\right)}$ multiplied by complicated factors consisting of trigonometric and hyperbolic functions. Both momentum and position appear together with the ratio $h/\hbar$ as arguments of these trigonometric and hyperbolic functions. 
As long as $h\ll\hbar$, there are no signs of any sub-Planck structures. These structures appear for sub-Planck values of position and momentum when $h \geq \hbar$.

As already mentioned, the superposition of the four coherent states is some generalization of so-called Schr\"odinger cat states. 
Therefore, in our case it is also possible to generate cat states from any pair of $N$ and $S$ or $E$ and $W$ states. 
Also, the Schr\"odinger equation corresponding to any component in~(\ref{compass-gen}) is a finite-difference equation with $h$ the step of the finite difference. 
Then one can take it equal to the Compton wavelength $\lambdabar=\hbar/mc$, which will allow one to explore the relation of the interference terms with relativistic corrections~\cite{jafarov,wall}. 

\section*{Acknowledgement}

This research was supported by project P6/02 of the Interuniversity Attraction Poles Programme (Belgian State -- 
Belgian Science Policy). E.I.\ Jafarov acknowledges the support of a Postdoc Fellowship within this programme.





\bibliographystyle{elsarticle-num}

\begin{thebibliography}{00}

\bibitem{zurek-n} W.H.~Zurek, Nature {\bf 412} (2001) 712-717.
\bibitem{schleich} W.~Schleich, M.~Pernigo, and F.L.~Kien, Phys. Rev. A {\bf 44} (1991) 2172-2187.
\bibitem{yurke} B.~Yurke and D.~Stoler, Phys. Rev. Lett. {\bf 57} (1986) 13-16.
\bibitem{hach} E.E.~Hach~III and C.C.~Gerry, J. Mod. Opt. {\bf 39} (1992) 2501-2517.
\bibitem{lynch} R.~Lynch, Phys. Rev. A {\bf 49} (1994) 2800-2805.
\bibitem{souza} A.L.~de~Souza~Silva, S.S.~Mizrahi and V.V.~Dodonov, J. Russ. Laser Res. {\bf 22} (2001) 534-544.
\bibitem{jacquod} Ph.~Jacquod, I.~Adagideli, and C.W.J.~Beenakker, Phys. Rev. Lett. {\bf 89} (2002) 154103.
\bibitem{toscano} F.~Toscano, D.A.R.~Dalvit, L.~Davidovich, and W.H.~Zurek, Phys. Rev. A {\bf 73} (2006) 023803.
\bibitem{banerji} J.~Banerji, Contemp. Phys. {\bf 48} (2007) 157-171.
\bibitem{scott} A.J.~Scott, and C.M.~Caves, Ann. Phys. {\bf 323} (2008) 2685-2708.
\bibitem{bhatt} J.R.~Bhatt, P.K.~Panigrahi, and M.~Vyas, Phys. Rev. A {\bf 78} (2008) 034101.
\bibitem{ghosh} S.~Ghosh, U.~Roy, C.~Genes, and D.~Vitali, Phys. Rev. A {\bf 79} (2009) 052104.
\bibitem{austin} D.R.~Austin, T.~Witting, A.S.~Wyatt, and I.A.~Walmsley, Opt. Commun. {\bf 283} (2010) 855-859.
\bibitem{jafarov} E.I.~Jafarov, S.~Lievens, S.M.~Nagiyev, and J. Van der Jeugt, J. Phys. A {\bf 40} (2007) 5427-5441.
\bibitem{atakishiyev} N.M.~Atakishiyev, and S.M.~Nagiyev, J. Phys. A {\bf 27} (1994) L611-L615.
\bibitem{ismail} M.E.H.~Ismail, C.F.~Dunkl, and R.~Wong, \textit{Special Functions}, World Scientific, Singapore, 2000.
\bibitem{gasper} G.\ Gasper and M.\ Rahman, \textit{Basic Hypergeometric Series}, Cambridge University Press, Cambridge, 1990.
\bibitem{koekoek} R.~Koekoek, and R. Swarttouw, \textit{The Askey-scheme of hypergeometric orthogonal polynomials and its $q$-analogue}, Delft University of Technology Report No 98-17, 1998.
\bibitem{wigner} E.P.~Wigner, Phys. Rev. {\bf 40} (1932) 749-759.
\bibitem{wall} F.T.~Wall, Proc. Natl Acad. Sci. USA {\bf 83} (1986) 5360-5363.

\end{thebibliography}



\end{document}